# An Investigation of the Factors Affecting the Transition of Lightning Surges through Underground Cables and Sub-Transmission Substations: A case study

Shahab Aref, Alireza Sedighi Anaraki, *Member, IEEE*

*Abstract*-- Lightning is a natural phenomenon that takes place all around the world. The enormous amount of energy released by lightning is extremely ruinous, so some suitable protection schemes are required for power systems to minimize its destructive effects. Given that the lightning surge passes through the transmission line, the destruction caused by lightning would not always happen at the strike point. Hence, the failure could take place at a point far from the striking point. This paper investigates factors affecting the transition of lightning surges through underground cables. These cables are utilized to supply a sub-transmission substation whose repeated cable failures have imposed enormous costs to regional electric distribution companies. To obtain reliable results, the highly capable software of EMTP was employed to calculate transient phenomena. In addition, various scenarios were simulated to identify the effects of some factors, including the effectiveness of surge arresters, bus-coupler switches, the grounding resistance of sheaths, and the like, on the susceptibility of the studied substation. Based on simulation results, the effects of the grounding resistance of cable sheaths at terminations and along the cable are significant.

*Index Terms*— Lightning, Underground Cables, Cross-Bonding, EMTP, Sub-Transmission Substation, Overvoltage, Bus-Coupler Switches

## I. INTRODUCTION

The overvoltage directed by lightning strikes to power system devices can lead to failures and massive outages. In fact, many parts of a power system are prone to undergo this sort of hazardous overvoltage. Transmission substations contribute to electricity transmission significantly. Lightning strikes hitting substation equipment can lead to a great blackout and subsequent huge costs. Due to the numerous protections of a substation, it is logical that the probability of direct lightning to its sensitive equipment be significantly lower than the strike to other equipment, such as overhead transmission lines [1]. In fact, a lightning strike hitting a transmission line near a substation is a potential cause of the failure in that substation, for lightning surges are able to travel alongside the transmission line and lead to a failure somewhere far away from their point of impact. The transition of such travelling waves is influenced by some factors, such as the grounding resistance of apparatuses, soil resistivity, the installation of surge arresters, and the like [2-5]. In the same vein, the corona phenomenon is a pivotal factor which could not be disregarded in certain situations [6].

The structures utilized to absorb renewable energies, such as wind and solar energies, are vulnerable to lightning strikes; thus, the proper implementation of protection schemes, such as the sheath grounding method, is of vital importance [7]. It is possible to predict the risk of faults caused by lightning in wind turbines in a specific region using the past lightning information of that region and a statistical machine learning approach [8]. Not having been protected adequately, thermal power plants would likewise be defenseless against lightning strikes. Thus, the suitability of their protection schemes, such as the grounding system, are required to be investigated [9]. The widespread use of HVDC lines, due to their unique benefits, requires special modelling methods and protection algorithms to withstand harmful lightning surges [10, 11].

There is no power system equipment whose influence on the simulation results of lightning strikes could be disregarded. Therefore, the accurate simulation of each instrument of the power system is effective in obtaining dependable results [12]. Successive cable failures in Montazer Qaem substation whose reasons are unknown has led to enormous costs for consumers and Yazd electrical distribution company. This paper aims to demonstrate the vulnerability of this sub-transmission substation to determine if there is a direct correlation between its successive cable failures and lightning strikes.

System configurations have been presented in section II. Appropriate models for simulating power systems are described in section III. Section IV is dedicated to simulations and respective results, under different scenarios. In the end, the conclusions of this paper are presented in section V.

## II. SYSTEM CONFIGURATIONS

A simulated power system consists of two sub-transmission substations connected by two 63-kV underground cables. The length of these underground cables is 7.5 km, which are employed to supply Montazer Qaem sub-transmission

The authors are with Department of Electrical Engineering, Yazd University, Iran (e-mail: shahabaref@stu.yazd.ac.ir; sedighi@yazd.ac.ir).

substation. These cables come from Daneshgah sub-transmission substation where two overhead lines turn into underground cables to feed Montazer Qaem substation. One of these overhead lines supplying Montazer Qaem substation comes from Yazd1 400-kV air-insulated substation. The aforementioned 400-kV substation has 16 (63-kV) feeders, with one of which 5.2 km long being used to supply Montazer Qaem substation. The other overhead line, being 21.8 km long, comes from Darvazeh Qoran sub-transmission substation. Fig. 1 represents the schematic diagram of the simulated power system.

## III. SYSTEM MODEL

The events experienced by power systems in terms of the frequency range are divided into four categories [13], with this frequency range being consisted of frequencies from 0.1 Hz to 50-MHz. Assigning appropriate models to each power system equipment for the simulation of any type of phenomenon, within its respective frequency range, is of vital importance to provide precise and reliable answers. Given that a lightning strike to a power system and its corresponding disturbances are classified as high-frequency phenomena, it is required to utilize high-frequency models commensurate with the type of the phenomenon. Various types of software are capable of simulating power systems, with one of which being Electromagnetic Transient Program (EMTP), which is highly specialized in simulating high-frequency phenomena in power networks. Accurate results from this software, which are almost indistinguishable from real experiences, have made it reliable. For each equipment, the proper model compatible with high-frequency events is extracted. Next, the models produced are imported into the software to see if the simulated power system is susceptible to disturbances originated from lightning strikes to the power system. The following sub-sections are assigned to the models described.

### A. Transmission line tower

There are disparate structures for towers in different situations in terms of their voltage level or being parallel with communication lines. In order to obtain precise results, towers are required to be modeled according to their structures. As already stated in [14], a simple distributed line model produces acceptable results for the simulation of 63-kV transmission line towers. The tower in this model is divided into four sections, with all of which having the same surge impedances. In fact, these sections are single-phase lossless distribute-parameter lines whose surge impedances are calculated according to (1).

$$Z = 60(\ln(\frac{H}{R}) - 1) \quad (1)$$

Where H is the tower's overall height, and R is defined as the equivalent radius of the tower calculated by (2).

$$R = \frac{r_1 h_2 + r_2 H + r_3 h_1}{2H} \quad (2)$$

Where $r_1$ is the radius of the tower tip, $r_2$ is the tower waist radius, $r_3$ is the tower bottom radius, $h_1$ is the height of the tower waist, and $h_2$ is distance between the tower tip and the tower waist [14]. Fig. 2 presents all aforementioned parameters upon considering that the tower is cone-shaped. According to the structure of a 63-kv transmission tower, the surge impedance of each section is equal to 130Ω.

### B. Tower footing resistance

The accurate simulation of footing resistance is of great importance, for reflections coming from the tower base return to the tower tip sooner than those coming from adjacent towers. If the simulation is done carelessly, the produced waveforms will not be reliable because inaccurate reflections are considered in the simulation. To take soil ionization into consideration, a non-linear resistor is considered as the grounding resistance of the towers. Being current-dependent, this resistor is explained through (3) [15].

$$R_T = \frac{R_0}{\sqrt{1 + \frac{I}{I_g}}} \quad (3)$$

Where $R_T$ is the tower footing resistance, $R_0$ is the tower footing resistance at low currents and low frequencies, $I_g$ is the current needed to initiate soil ionization, and I is the lighting current passing through the footing resistance [15]. The current needed to initiate soil ionization depends on some factors whose relation is presented in (4).

$$I_g = \frac{1}{2\pi} \cdot \frac{E_0 \rho}{R_0^2} \quad (4)$$

Where $R_0$ is soil resistivity (Ω.m), and $E_0$ is the soil ionization gradient (kV/m) [15]. For the simulation carried out in this paper, the amounts of $R_0$, $E_0$, and ρ are 10Ω, 400 kV/m, and 100 Ω.m, respectively.

### C. Insulator strings

A voltage-dependent flashover switch, parallel with a capacitor, is presented for each insulator string. The capacitor is utilized to represent the coupling effect of the phase conductor on the tower structure. The leader progression method (LPM) is adopted to represent the physical aspects of the discharge mechanism, as stated by (5) [16] as follows:

$$\frac{dl}{dt} = k_1 v(t)(\frac{v(t)}{g-1} - E_{10}) \quad (5)$$

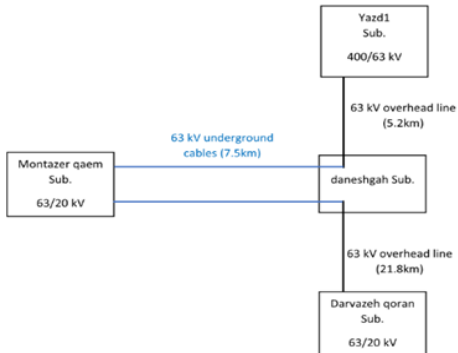

Fig. 1. Schematic of simulated power system

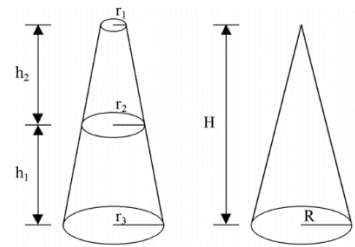

Fig. 2. Tower equivalent radius [14]

Where l is the leader length (m), g is the gap length (m), v(t) is the voltage across the gap (kV), and $k_1$ and $E_{10}$ are constant values equal to $2\times 10^{-7}$ m²/v²s and 400 kV/m, respectively. The magnitude of the paralleled capacitor is 16 PF for each insulator string.

### D. Overhead transmission line

Due to the high frequency of lightning strikes, a 63-kV overhead transmission line is simulated by an un-transposed frequency-dependent model for each line span. The skin effect is regarded for the calculation process of the line parameters. Furthermore, the frequency at which these parameters are calculated is 500 kHz [17].

The geometrical position, DC resistance, and the dimensions of all conductors are the information required for creating a line model. The magnitude of DC resistance and the external radius of the utilized wire as the phase conductor are 0.1576 Ω/km and 1.953 cm, respectively. The aforementioned parameters are 1.92 Ω/km and 0.837 cm for the guard conductor, respectively. Five line spans are simulated on both sides of the point at which lightning strikes, in which reflections from adjacent towers are vital to be considered. The length of each line span is equal to 280 m, yet the length of the last span is 20 km so as to prevent reflections from the open end at the last tower. The overhead line section through which lighting sweeps is not long enough to consider the corona effects [18].

### E. Underground cable

A proper model is required to be produced for underground cables, which is covering the parameter changes of the cables commensurate with frequency variations. A frequency-dependent distributed parameter model is utilized in this paper to obtain reliable results. [19]. FDQ is the model through which the frequency dependency of cable parameters is considered in EMTP, which stands for the frequency-dependent Q matrix. Prerequisites for creating the FDQ model are the cables' geometrical and structural data, such as the internal and external radius of conductors and insulators, their electrical parameters, the burial depth of the cables, and the way they are laid out. The two simulated sub-transmission substations of Montazer Qaem and Daneshgah are connected by a double circuit underground cable 7.5 km long. The cross-section of these cables is 300 mm² with cross-linked polyethylene (XLPE) utilized as the dielectric to separate the copper conductor from the aluminum sheath. The insulation thickness and overall diameter of the cables are 11 mm and 66 mm, respectively. These cables are buried 120 cm deep in the ground, and the distance between them is twice as thick as the the size of the cable diameter. To protect the cables from voltage spikes, they are divided into 700-m sections, which are cross-bonded and grounded by means of surge arresters at the end of each section. Other information necessary for simulating the cables are presented in [19].

### F. Surge arresters

Surge arresters are widely used to protect significant and valuable equipment in the power network from harmful events. The frequency-dependent surge arrester model presented by IEEE, which is employed in this paper, is of high accuracy. This model is composed of two time-independent nonlinear resistors ($A_0$ and $A_1$), a couple of linear inductors ($L_0$ and $L_1$), two linear resistors ($R_0$ and $R_1$), and a capacitor (C) [20]. Fig. 3 shows the surge arrester model. To compute the initial magnitude of each element, the height of the surge arrester and the number of ZnO parallel disks are needed to be specified. The simulated surge arresters have one parallel disk, with the height of the arresters being 0.641 m and 0.725 m for the voltage levels of 20 kV and 63 kV, respectively [20].

### G. Transformer

Surge transmission through the transformer is dependent on a number of factors, such as the voltage level, the turn ratio, as well as the electrostatic and electromagnetic couplings of the windings [21]. In fact, the capacitors within the transformer play a pivotal role in surge transition through it. These capacitances are commensurate with the physical structure of the transformer and cannot be acquired easily. A very detailed model consists of all inductances and capacitances within each turn of the windings and between every two adjacent winding turns. Fig. 4 shows the transient model for the transformer employed in this paper, in which $C_{bh}$ is the high-voltage bushing capacitance, $C_b$ is the secondary bushing capacitance, $C_{hg}$ is the distributed capacitance of the high-voltage windings, $C_{lg}$ is the distributed capacitance of the low-voltage winding, and $C_{hl}$ is the distributed capacitance between the high and low-voltage windings. Given that capacitances on each side are parallel, they can be simplified. After simplification, in accordance with the transformer's rated power which is 30 MVA, the high-voltage and low-voltage capacitances, and the capacitances between the high and low-voltage windings will be 4.2 nF, 20 nF, and 18 nF, respectively [21].

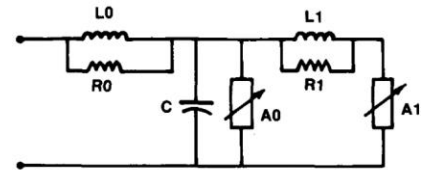

Fig. 3. The transient model of surge arrester [20].

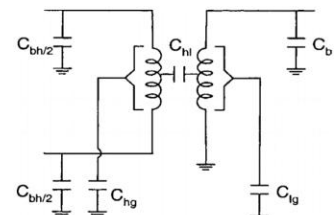

Fig. 4. Simplified transient model of transformer [21].

## H. Substation equipment

Transmission and sub-transmission substations consist of different equipment, such as circuit breakers, as well as current, potential, and capacitor voltage transformers, and the like. To simulate transient phenomena in the power system precisely, all these equipment are required to be modelled.

Proper models are extracted from [17, 22]. In accordance with the models for each equipment, simulation results from the studied system are presented in the appendix.

The lightning strike represented by CIGRE [16] has been employed in this paper. This concave lighting strike is shown in Fig. 5, in which $I_f$ is the crest current, $S_m$ is the maximum front steepness, $t_f$ is the equivalent front duration, and $t_h$ is the time-to-half value. Given that the discharged current tends to disperse in the atmosphere, 1500-Ω resistance is considered in parallel with the lighting current source [23]. Table I shows the magnitude of the parameters of the lightning strike.

## IV. SIMULATION RESULTS

In this section, the simulation results of the lightning strike hitting the last transmission tower in the vicinity of Daneshgah substation are presented, where overhead lines turn into underground cables. To assess the vulnerability of Montazer Qaem substation to these factors, several influential factors are investigated.

Lightning strikes hitting the shield wire of the tower are simulated as the first scenario. According to the simulation results, the transformer of the Montazer Qaem substation experiences minor disturbances, in that surge arresters at Daneshgah and Montazer Qaem substations dampen the sharp harmful overvoltage. The shielding failure is considered as the next scenario in which lightning strikes phase A.

Because of the presence of surge arresters in sub-transmission substations as well as the long distance between the two substations, lightning disturbances are dampened, and the maximum voltage on the high-voltage side of Montazer Qaem transformer reaches 110 kV. To intensify the lightning impact, the maximum current of lightning increases to 90 kA. However, the voltage magnitude of the transformer increases only by 7 kV, which shows the remarkable effectiveness of surge arresters in stabilizing the power system. Fig. 6 shows the voltage waveforms of the two scenarios.

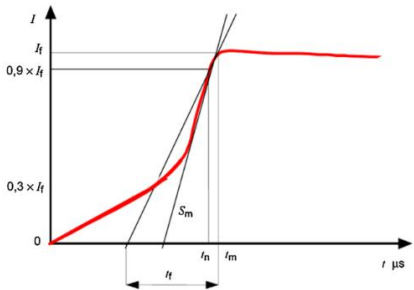

Fig. 5. Lightning strike shape of CIGRE [24].

Table I
Parameters of CIGRE lightning strike

| $I_f$ (kA) | $S_m$ (kA/μs) | $t_f$ (μs) | $t_h$ (μs) |
|---|---|---|---|
| 31 | 26 | 3 | 75 |

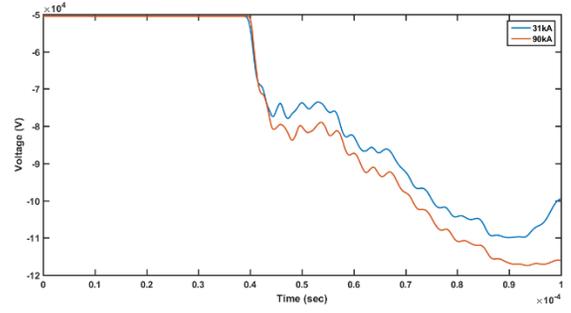

Fig. 6. Transformer's primary side voltage when lightning strikes phase A ($I_f$ = 31 kA, 90 kA)

The sheaths of the underground cables are connected to each other and grounded at the beginning of the cables at Daneshgah substation and at the end of the cables at Montazer Qaem substation.

The sheaths are grounded at Montazer Qaem and Daneshgah substations by 10-Ω and 20-Ω resistances, respectively. The influence of the resistance using which the sheaths are grounded is investigated as the next scenario. The magnitude of these two resistances varies from 0 Ω to 90 Ω simultaneously. To obtain satisfactory results, surge arresters are disconnected at both substations. According to Fig. 7, an increase in sheath resistance leads to a decrease in the voltage on the primary side of the transformer. The peak magnitude of current which flows from sheaths to ground at Daneshgah substation when sheaths are directly grounded is 10239 A and it is 1667 A when sheaths are grounded by a 90 Ω resistance. The peak magnitude of voltage at this substation when cable sheaths are directly grounded and by a 90 Ω resistance are 419 kV and 518 kV, respectively. However, the implementation of this resistance culminates in a less voltage peak at Montazer Qaem substation than being directly grounded. The joints at which the cable sheaths are cross-bonded are numbered from one to 10. The first joint is near Daneshgah substation, and the last or the tenth joint is in the vicinity of Montazer Qaem substation. The voltage waveform of Daneshgah substation, joint 1, joint 4, joint 7, joint 10 and transformer of Montazer Qaem substation when sheaths at the beginning and end of cable are directly and by a 90 Ω resistance grounded are shown in Figs. 8 and 9. The maximum magnitude of the lightning current is superseded by 90 kA, with the simulation reiterated. Simulation results are presented in Fig. 10, with its trend being similar to Fig. 7, but with higher magnitude.

As already stated, sheaths are transposed and grounded by sheath voltage limiters (SVL) throughout the cable length. Another scenario concerning the effects of grounding resistance to which SVL is connected has been simulated. The resistance magnitude varies from 0 Ω to an infinite value (an open circuit). Fig. 11 shows the effect of resistance variations. The greatest voltage magnitude belongs to the situation where the resistance value is zero.

In fact, the first voltage peak occurs at about 45 μs, and the second one occurs at about 75 μs. According to Fig. 11, the trend of the voltage peak has been reversed from the first peak to the second one. The first voltage peak belongs to the short circuit, and the second voltage peak occurs at the open circuit.

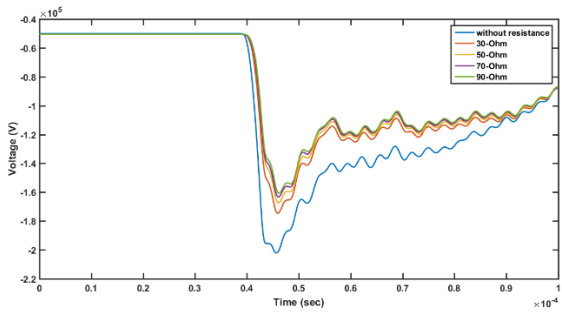

Fig. 7. Transformer's primary side voltage in accordance with variation of sheath grounding resistance at both sides of cable ($I_f$ = 31 kA).

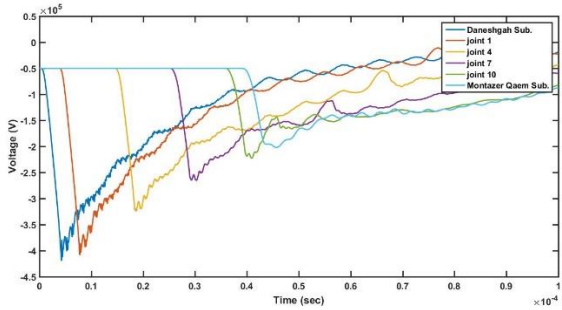

Fig. 8. Voltage waveform along cable (cable sheaths directly grounded)

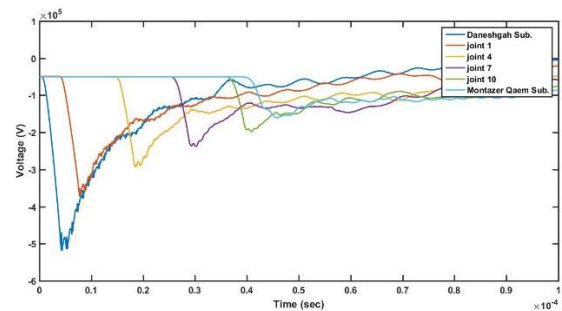

Fig. 9. Voltage waveform along cable (cable sheaths grounded by a 90 Ω resistance)

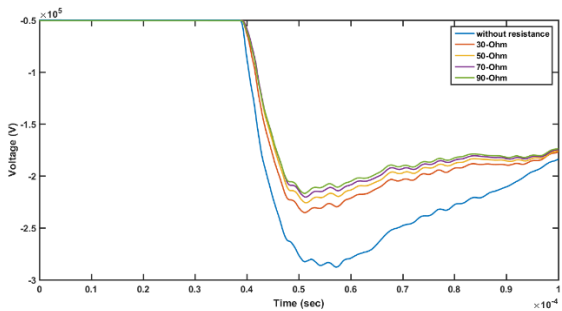

Fig. 10. Transformer's primary side voltage in accordance with variation of sheath grounding resistance at both sides of cable ($I_f$ = 90 kA).

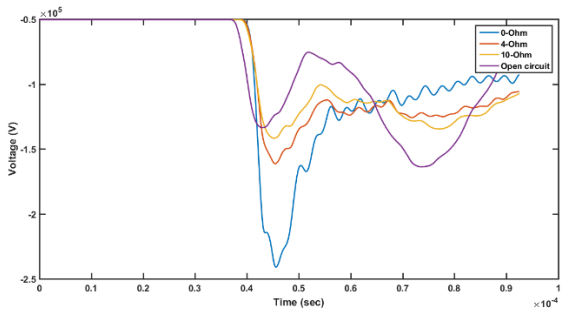

Fig. 11. The effect of grounding resistance in cross bonding points.

In fact, reflections from different points in the power system lead to this inverse trend.

In another simulation, SVLs are eliminated to examine their effectiveness. Fig. 12 shows the results of this simulation according to which there is a trend similar to that of Fig. 11 but with higher magnitude. According to Fig. 12, the absence of SVLs leads to the higher peak voltage of 7 kV.

Montazer Qaem substation has two transformers connected by a bus-coupler switch. The lightning surge spreads in the substation by this bus-coupler switch. In fact, phase A of the first circuit is struck with lightning during the simulation. Next, this surge passes through the first circuit of the underground cable in the substation and affects the second circuit. The impacts of the bus-coupler switch on the travelling waves and the voltage of transformers are investigated in the next scenario. Fig. 13 shows the voltage waveforms of the transformer of the first circuit when the bus-coupler switch is considered to be closed and open. According to this figure, if this switch remains open, the lightning energy will not spread through the substation. Hence, the first transformer will undergo significant overvoltage.

Fig. 14 shows the voltage waveform of the second transformer for this scenario. The closed bus-coupler switch makes the second transformer undergo voltage disturbances the same as the first transformer; however, the open switch culminates in minor disturbances.

The effects of the grounding resistance of the sheaths on both sides of the cables on the traveling waves are investigated as well. If the bus-coupler switch is closed, a considerable amount of the lightning energy will come back to Daneshgah substation through the second circuit of the underground cables. Given that Daneshgah substation is the connection point between the underground cable and the overhead line, the magnitude of the surge coming from the underground cable will increase in accordance with different surge impedances [25]. Although voltage increase in Daneshgah substation is inevitable, the effects of the grounding resistance of the sheaths in that substation are significant. Fig. 15 shows the voltage waveform of phase A of the second circuit at Daneshgah substation and four joints, including joint 1, joint 4, joint 7, and joint 10. According to Fig. 15, the voltage waveform decreases continuously until it arrives at Daneshgah substation where the voltage increases with the surge impedance of the cable being less than that of the overhead line.

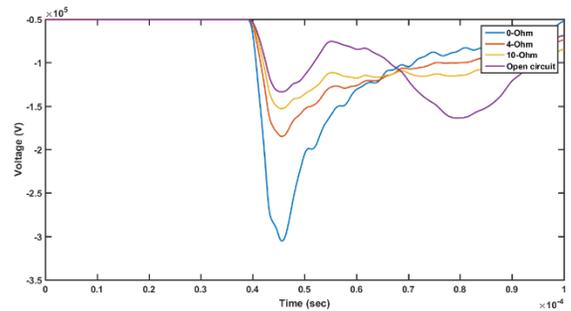

Fig. 12. The effect of grounding resistance in cross bonding points (SVLs have eliminated).

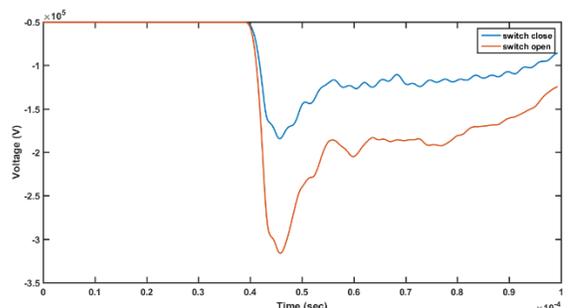

Fig. 13. The effect of bus-coupler switch on the voltage of first transformer.

The grounding resistances of the sheaths on both sides of the cables were eliminated, and the simulation was repeated. According to Fig. 16, the elimination of these resistances leads to the voltage increment of about 8 kV at Daneshgah substation.

The last simulation is dedicated to investigating the effectiveness of the sheaths in inducing voltage to nearby phases. Two situations are considered in this simulation, with one of which being normal operating conditions, and the other being the elimination of cross-bonding which indicates the sheath of each cable will not be transposed. According to Fig. 17, the simulation results of normal operating conditions imply that upon the passage of the surge through the cable, the voltage induced to phase C increases. Due to cross-bonding, the voltage induced to the sheath of phase A will be induced to all other phases. According to Fig.18, the voltage induced to phase C reduces significantly when the sheaths are not cross-bonded.

## V. CONCLUSIONS

Lightning is a natural phenomenon, and when it strikes an unprotected power system, it leads to enormous damage. It is of vital importance to predict hazardous situations where power networks are vulnerable. There are some specialized types of software capable of simulating transient phenomena, with one of which being EMTP which is employed in this paper. After simulating all equipment in EMTP, the vulnerability of Montazer Qaem substation to lightning strikes was investigated. According to the simulation results, if the surge arresters located in Daneshgah and Montazer Qaem substations work properly, Montazer Qaem substation will not suffer any damage. Underground cables with the length of 7.5 km had similar meaningful effects on the reduction of the transients. Due to the simulation results, the selection of appropriate grounding resistance for the sheaths is of vital importance.

## VI. APPENDIX

Fig. A-I shows the simulated power system in the EMTP software.

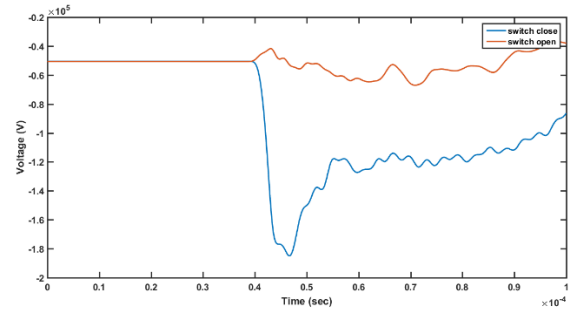
Fig. 14. The effect of bus-coupler switch on the voltage of second transformer.

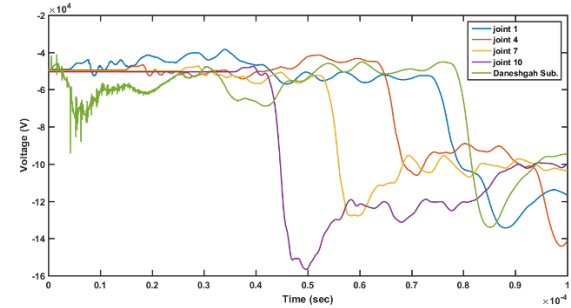
Fig. 15. Voltage waveform in the second circuit of underground cable.

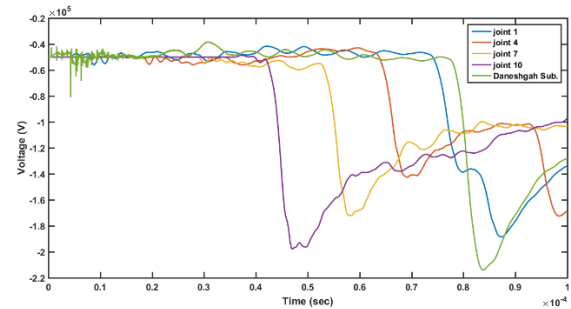
Fig. 16. Voltage waveform in the second circuit of underground cable (sheaths resistances eliminated).

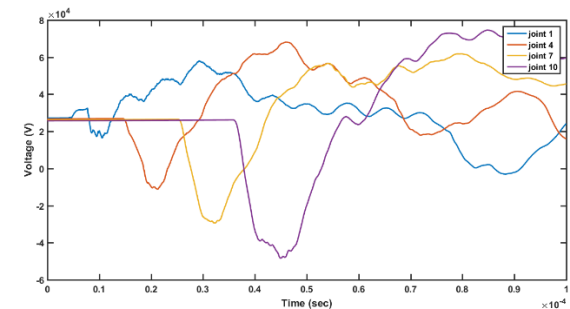
Fig. 17. Induced voltage on phase C.

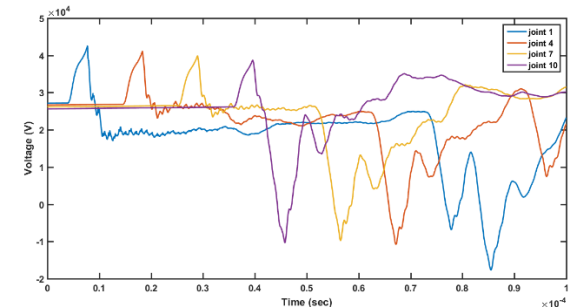
Fig. 18. Induced voltage on phase C (sheaths have not transposed).

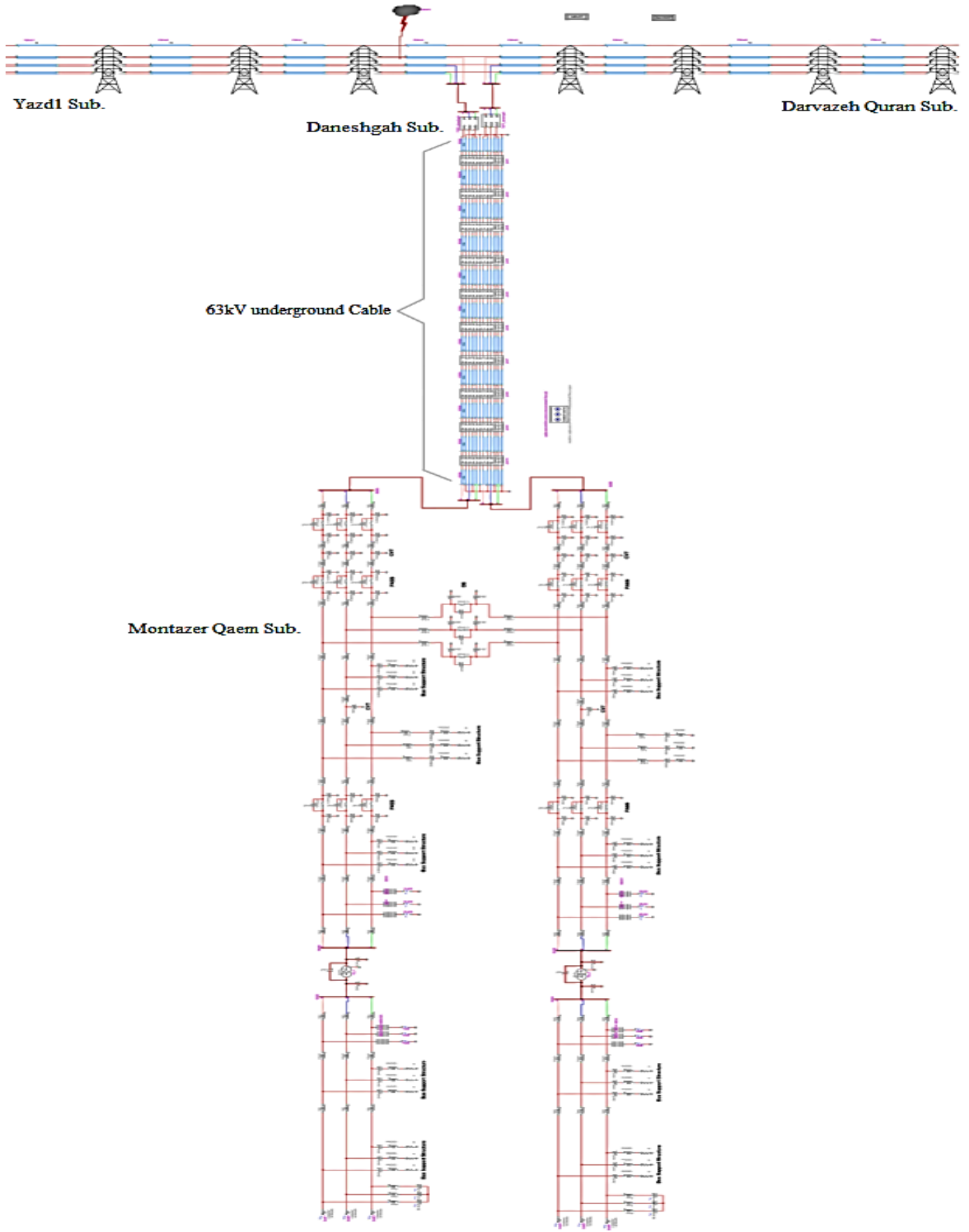

Fig. A-I. Simulated power system in EMTP